\documentclass[3p,times]{elsarticle}

\usepackage{ecrc}

\volume{00}

\firstpage{1}

\journalname{Nuclear Physics A}

\runauth{}

\jid{nupha}

\usepackage{amssymb}

\usepackage{lineno}

\usepackage[figuresright]{rotating}

\usepackage{graphicx}
\usepackage{caption}
\usepackage{subcaption}

\begin{document}

\begin{frontmatter}



\dochead{}

\title{$\Upsilon$ Measurements by the PHENIX Collaboration}


\author{Shawn Whitaker(for the PHENIX Collaboration)}

\address{Iowa State University, Ames, IA, USA}

\begin{abstract}

PHENIX has measured $\Upsilon$ production in p+p, d+Au and Au+Au collisions at $\sqrt{s_{NN}}$=200 GeV. A baseline cross-section measurement has been made in p+p collisions.  Preliminary studies of d+Au collisions found no suppression in the backward direction, but suppression was observed in the forward direction indicating the presence of cold nuclear matter effects.  In Au+Au collisions preliminary studies of $\Upsilon$ production at mid-rapidity found hints of suppression.


\end{abstract}


\end{frontmatter}



\section{Introduction}
\label{sec:intro}

\subsection{Motivation}

Quarkonia suppression has long been thought to be a signature of the creation of the quark-gluon plasma\cite{matsui-satz}.  By measuring the suppression of various quarkonia states it is possible to put constraints on the temperature of the QGP produced in heavy ion collisions due to the different binding energies of the states. However, given that the temperature of the medium produced at RHIC has been measured using direct photons at PHENIX it seems more appropriate to leverage the varying sizes of the quarkonia bound states to further constrain the temperature dependent Debye screening length of the medium.

\subsection{PHENIX Detector}

The PHENIX experiment has two sets of detectors for different rapidity ranges: the central arms used for measuring $\Upsilon$s decaying to di-electrons at mid-rapidity ($|y| \le 0.35$) and the muon arms used for measuring $\Upsilon$s decaying to di-muons at forward and backward rapidities ($1.2\le|y|\le2.2$).   The detectors used in the central arms were the drift chamber for momentum reconstruction, the Ring Imaging Cherenkov detector (RICH) for electron identification and finally the lead glass and lead scintillators used for electromagnetic calorimetry and further electron identification using an E/p measurement.  The muon arms are made up of two primary detectors, the muon identifier which is made up of alternating layers of iron and larocci tubes and the muon tracker which consist of cathode strip tracking chambers in a magnetic field.

\begin{figure}[ht]
\begin{subfigure}[b]{0.45\textwidth}
\raisebox{0.5cm}{\includegraphics[width=\textwidth]{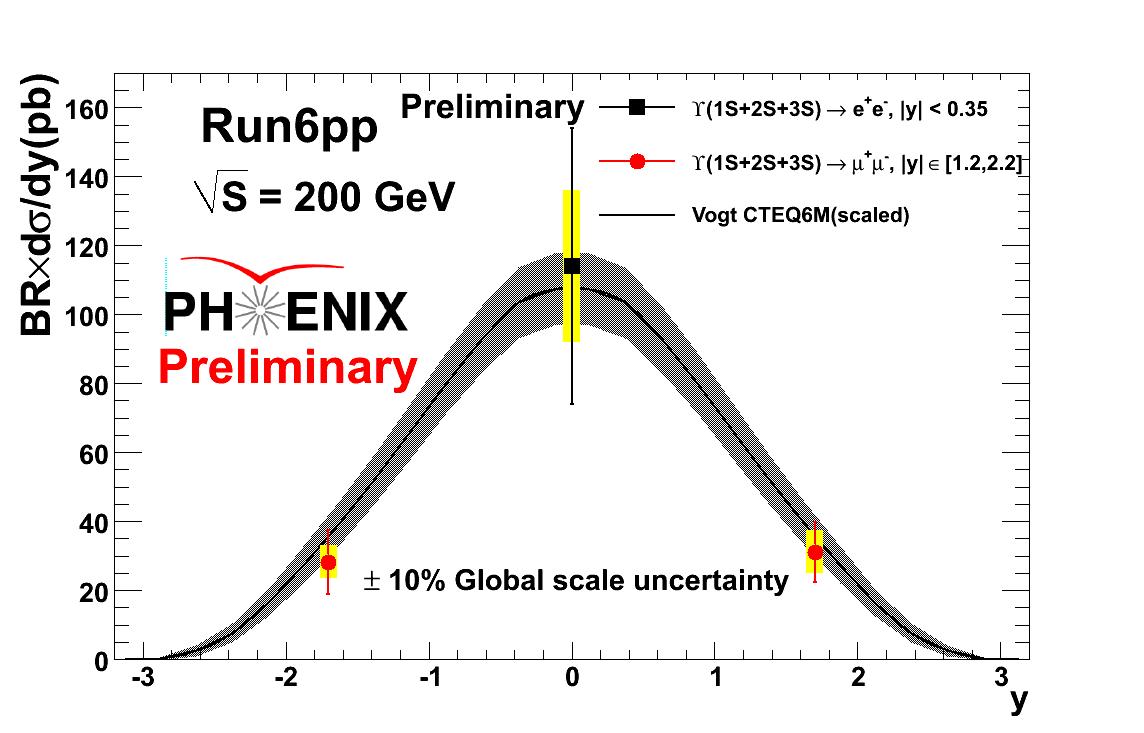}}
\caption{}
\label{fig:pp rap dep cross section}
\end{subfigure}
~
\begin{subfigure}[b]{0.45\textwidth}
\includegraphics[width=\textwidth]{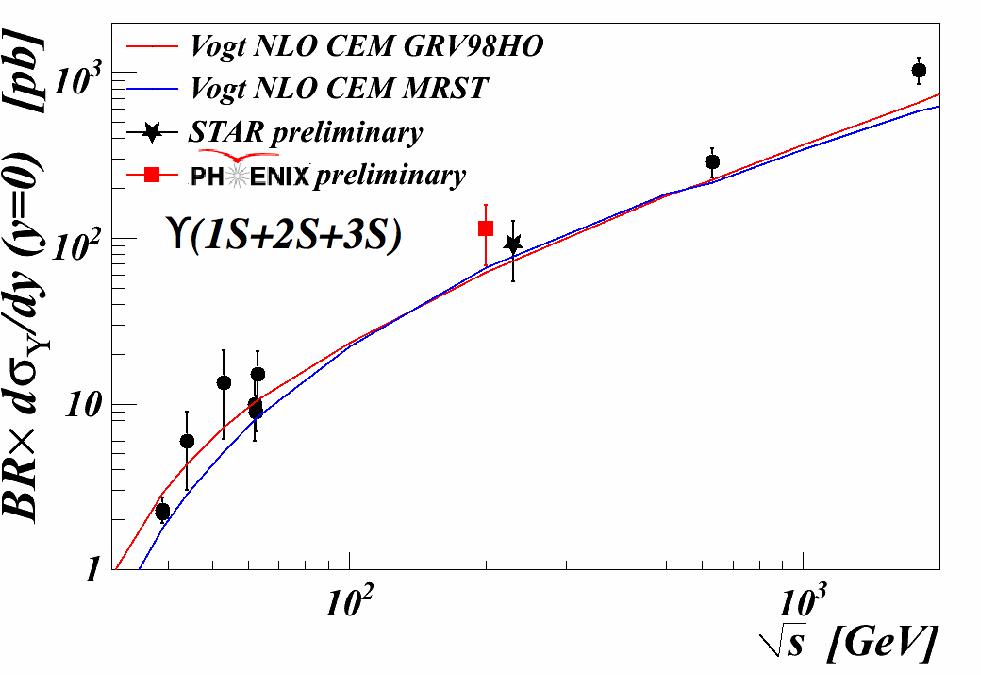}
\caption{}
\label{fig:pp energy dep cross section}
\end{subfigure}
\caption{Measurements of $\Upsilon$ from 2006 p+p PHENIX data. (a) $\Upsilon$ cross section as a function of rapidity.  The black band corresponds to a R. Vogt theoretical calculation, scaled to match the data.  (b) $\Upsilon$ mid-rapidity cross section plotted as a function of collision energy. PHENIX data is at $\sqrt{s}=200 GeV$ consistent with global trends.} 
\label{fig:run6}
\end{figure}

\section{Method}

Invariant mass spectra were created from opposite signed di-electrons and di-muons and $\Upsilon$ candidates were selected from those pairs with an invariant mass between 8.5 GeV and 11.5 GeV.  Once the sample of candidates were selected the various backgrounds needed to be accounted for.  These backgrounds were from two sources; combinatorial background from the random association of lepton pairs, and correlated backgrounds from physical processes such as Drell-Yan and the semi-leptonic decays of open charm and open bottom.

The combinatorial background was removed by creating an invariant mass spectra for same sign lepton pairs and subtracting that distribution from the opposite signed distribution.

Correlated background is studied simulating the shapes of the Drell-Yan and open charm and bottom contributions. The Drell-Yan calculation, while the open bottom and open charm shapes are fitted to the data, having their normalization as a free parameter of the fit. The Upsilon yields are extracted by counting the number of opposite signed pairs in the mass window 8.5-11.5 GeV, after the subtraction of the previously evaluated backgrounds.

\section{Results}

In this section I will summarize the $\Upsilon$ results from PHENIX. I will begin with the p+p measurements at forward, backward and mid-rapidity which are used as a baseline for d+Au and Au+Au analysis. I will then discuss the nuclear modification factors at forward and backward rapidities in d+Au and at mid-rapidity in Au+Au collisions.

\subsection{p+p}

PHENIX measurements of the $\Upsilon$ cross section at $\sqrt{s}=200 GeV$ were made at forward, backward, and mid-rapidities.  These were shown in Fig. \ref{fig:pp rap dep cross section} together along with a scaled rapidity dependent cross section calculation from R. Vogt based on a CTEQ6M model.  In addition the mid-rapidity cross section measurement was plotted as a function of energy in Fig. \ref{fig:pp energy dep cross section} and shown with measurements from other experiments as well as a couple of NLO color evaporation model curves from R. Vogt.  The mid-rapidity measurement from PHENIX was shown to be in good agreement with global trends and theoretical models.  These p+p measurements form a strong foundation as we move forward and investigate the effects of both hot and cold nuclear matter (CNM) on $\Upsilon$ production.

\subsection{d+Au}

\begin{figure}
\includegraphics[width=\textwidth]{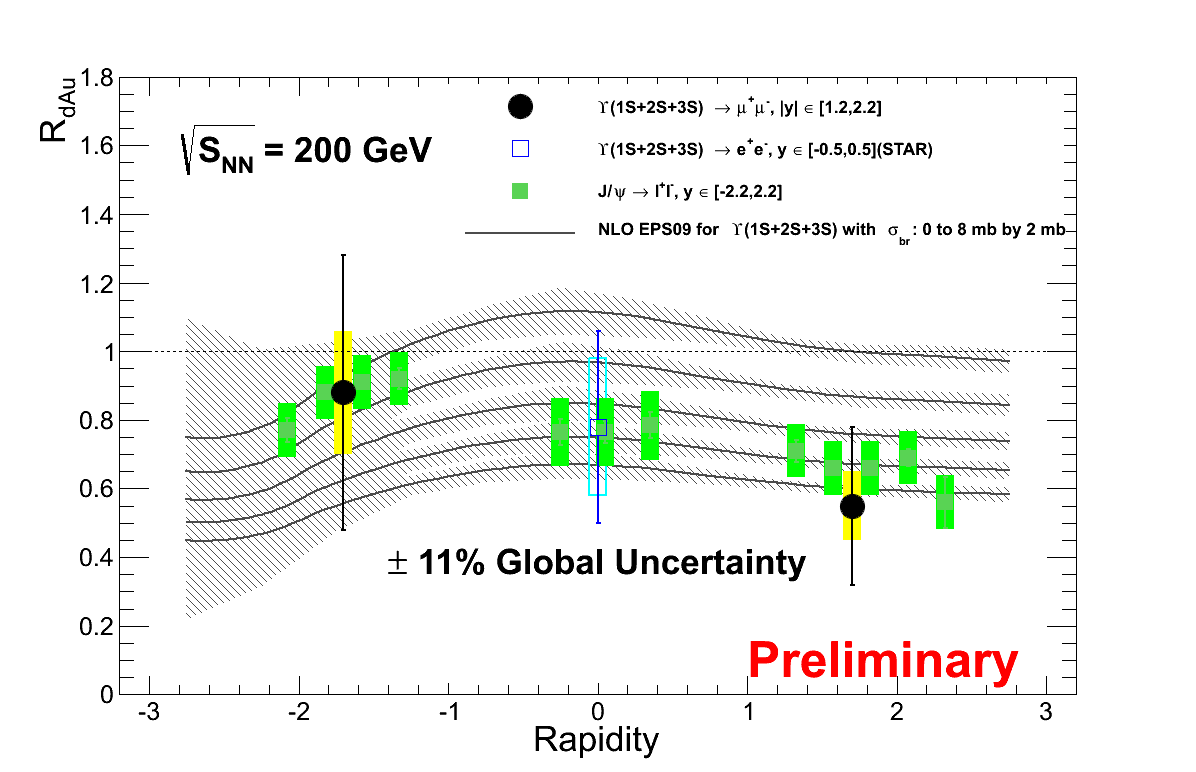}
\caption{$\Upsilon$  R$_{dAu}$ plotted as a function of rapidity with data from STAR at mid-rapidity, comparison to PHENIX J/$\psi$ R$_{dA}$(green points) measurements as well as a NLO EPS09 calculation with various breakup cross sections (black lines).}
\label{fig:RdA}
\end{figure}

With a p+p baseline in hand it is time to look into the effects of the presence of nuclear matter on $\Upsilon$ production. To separate CNM effects from effects due to the presence of the QGP it is necessary to measure $\Upsilon$s in d+Au collisions to create a CNM baseline.  Preliminary measurements from PHENIX in the forward and backwards directions were shown and compared to a STAR measurement at mid-rapidity, PHENIX J/$\psi$ results and a NLO EPS09 calculation with a breakup cross section varying from 0 mb to 8 mb all of which is shown in Fig. \ref{fig:RdA}. No suppression was seen at backward rapidity.  Suppression was observed at forward rapidity indicative of CNM effects.  The trend of increasing suppression as we move forward in rapidity is supported by the comparison with STAR data.  In addition these preliminary measurements were compared to PHENIX J/$\psi$ results and were found to be in remarkable agreement.  When compared to a NLO EPS09 calculation with varying breakup cross sections it is clear that the theory cannot as of yet reproduce the data across all rapidities.

\subsection{Au+Au}

The nuclear modification factor, R$_{AuAu}(\Upsilon)$ from the 2010 minimum bias data sample was calculated based on the relative yields of $\Upsilon$ with J/$\psi$ correcting for the relative acceptance and occupancy effects found from Monte Carlo and correcting the J/$\psi$ yield for the known suppression found in Au+Au collisions as shown in Eq. \ref{eq:double ratio}.  

\begin{equation}
R_{AuAu}(\Upsilon)=\frac{[N(\Upsilon)/N(J/\psi)]_{AuAu}}{[N(\Upsilon)/N(J/\psi)]_{pp}}\times \varepsilon_{rel} \times R_{AuAu}(J/\psi)
\label{eq:double ratio}
\end{equation}

\begin{figure}
\begin{subfigure}[b]{0.45\textwidth}
\includegraphics[width=\textwidth]{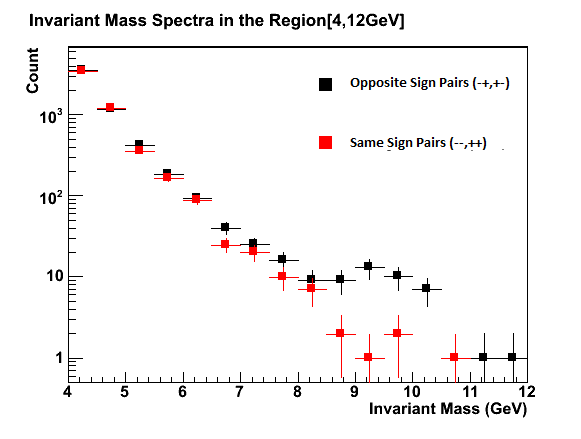}
\caption{}
\label{fig:mass spectra}
\end{subfigure}
~
\begin{subfigure}[b]{0.45\textwidth}
\raisebox{0.4cm}{\includegraphics[width=\textwidth]{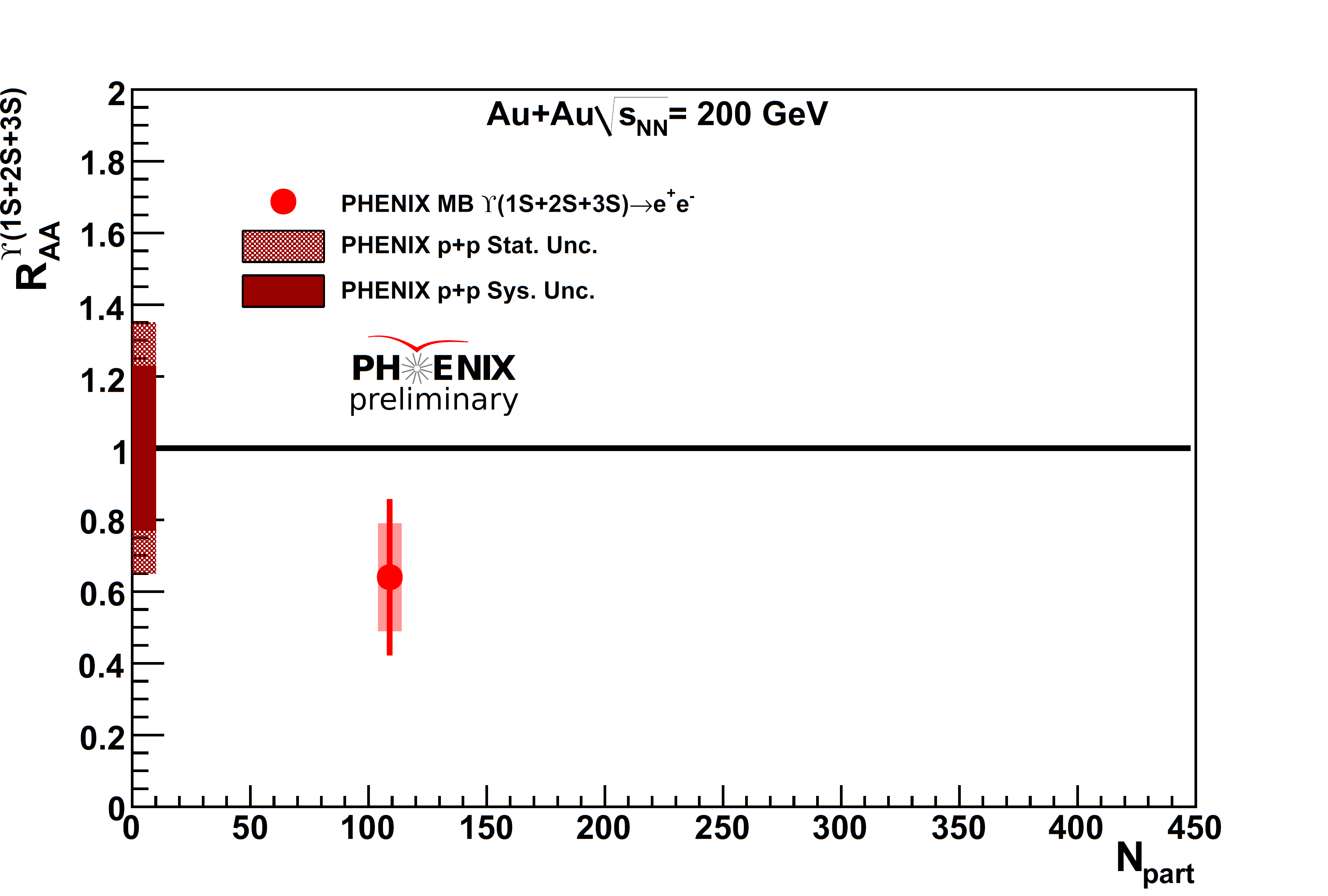}}
\caption{}
\label{fig:RAA}
\end{subfigure}
\caption{$\Upsilon$ measurements from 2010 Au+Au PHENIX data. (a) Invariant mass spectra for the 2010 Au+Au mid-rapidity minimum bias data at PHENIX.  (b) R$_{AuAu}(\Upsilon)$ measurement for minimum bias data sample plotted as a function of $N_{part}$.}
\label{fig:run10}
\end{figure}

The invariant mass spectra found in the 2010 Au+Au data are shown in Fig. \ref{fig:mass spectra} where a significant excess of the opposite signed electron pairs can be seen over the like sign electron pair spectra.  A preliminary measurement of R$_{AuAu}(\Upsilon)$ was constructed from the yields obtained from these spectra, the yields of $\Upsilon$ and J/$\psi$ from previous p+p measurements and a previously published\cite{jpsi_raa} R$_{AuAu}(J/\psi)$ measurement.  The result is shown in Fig. \ref{fig:RAA} as a function of N$_{part}$ and was found to be in good agreement with data shown by STAR at Quark Matter 2011\cite{starQM11}.

\section{Conclusion}

PHENIX has measured $\Upsilon$ cross sections at a wide range of rapidities in p+p collisions, along with preliminary measurements of R$_{dAu}$ at forward and backward rapidities in d+Au collisions and R$_{AuAu}$ at mid-rapidity in Au+Au collisions.  At the time of this presentation work was still being done on completing the mid-rapidity measurement from d+Au, and the forward and backward measurements in Au+Au.  With continual improvements of RHIC and upgrades at PHENIX these early measurements mark the beginning of a rich $\Upsilon$ program at PHENIX that will complement the very extensive J/$\psi$ results that have already been published.








\end{document}